# Nitrogen Substituted Polycyclic Aromatic Hydrocarbon As Capable Interstellar Infrared Spectrum Source Considering Astronomical Chemical Evolution Step To Biological Organic Purine And Adenine

NORIO OTA
Graduate School of Pure and Applied Sciences, University of Tsukuba,
1-1-1 Tenoudai Tsukuba-city 305-8571, Japan; n-otajitaku@nifty.com

In order to find out capable chemical evolution step from astronomically created organic in interstellar space to biological organic on the earth, infrared spectrum (IR) of nitrogen substituted carbon pentagon-hexagon coupled polycyclic aromatic hydrocarbon (N-PAH) was analyzed by the density functional theory. Ionization was modeled from neutral to tri-cation. Among one nitrogen and two nitrogen substituted N-PAH, we could find good examples showing similar IR behavior with astronomically well observed one as like $C_8H_6N_1$-b$^+$, $C_7H_5N_2$-bb$^+$, and $C_7H_5N_2$-ab$^{3+}$. We can imagine that such ionized N-PAH may be created in interstellar space by attacks of high energy nitrogen and photon. Whereas, in case of three and four nitrogen substituted cases as like $C_6H_4N_3^{n+}$ and $C_5H_3N_4^{n+}$ (n=0, 1, 2, and 3), there were no candidate showing similar behavior with observed one. Also, IR of typical biological organic with four and five nitrogen substituted one as like purine ($C_5H_4N_4$) and adenine ($C_5H_5N_5$) resulted no good similarity with observed one. By such theoretical comparison, one capable story of chemical evolution of PAH's in interstellar space was that one and two nitrogen substituted carbon pentagon-hexagon molecules may have a potential to be created in interstellar space, whereas more than three nitrogen substituted molecules including biological organic purine and adenine may not be synthesized in space, possibly on the earth.

## 1, INTRODUCTION

One of the important question in nature history is when, how, where basic organic material of living life was chemically prepared. This also means to study a possible chemical evolution step in the universe. In this paper, we supposed that simple polycyclic aromatic hydrocarbon (PAH) like benzene, naphthalene etc. were created in interstellar space by supernova and new star born shockwave. After that, high energy carbon, nitrogen, and photon attack and modulate such basic PAH to biological organic as like purine, adenine. As illustrated in Figure 1, some simple organics could be created in interstellar space, named "astronomical organic". On the other hand, we really know existence of biological organics on the earth. We like to know which chemical is the border between those two. This paper will compare density functional theory (DFT) based calculated infrared spectrum (IR) with astronomically well observed one varying number of substituted nitrogen. Nitrogen is an important atom playing key role in biological organics.

It is well known that interstellar PAH shows ubiquitous specific infrared (IR) spectrum from 3 to 20μm (Boersma et al. 2014). Recently, Tielens pointed out in his review (Tielens 2013) that void induced PAH may be one candidate. According to such previous expectations, we tried to calculate IR on one typical example of void coronene $C_{23}H_{12}^{++}$ (Ota 2014, 2015a, 2015b), which show an amazing result that this single molecule could almost reproduce a similar IR spectrum with astronomically well observed one. The current central concept to understand observed astronomical spectra is the decomposition method from the data base of many PAH's experimental and theoretical analysis (Boersma et al. 2013, 2014). In order to explain our unexpected result, naphthalene $C_{10}H_8$ and its void induced cation $C_9H_7^{n+}$ was analyzed in a previous note (Ota 2015c). In interstellar space, especially new star born area, high energy proton and photon attack PAH molecules to bring void induced PAH and its cation. Void induced carbon pentagon-hexagon molecule $C_9H_7^{n+}$ has intrinsic electric dipole moment by its asymmetrical geometry and may bring fairly good IR behavior. We enhanced such study from benzene $C_6H_6$ (single carbon ring) to naphthalene $C_{10}H_8$ (two rings), and 1H-phenalene $C_{13}H_9$ (three rings), which brought similar conclusion that asymmetrical carbon pentagon-hexagon PAH's were candidates (Ota 2015d). Here, we like to start from $C_9H_7$ and substitute carbon sites by nitrogen. As imaged in Figure 2, every carbon site of $C_9H_7$ may be attacked and substituted by high energy nitrogen. At the same time, high energy photon attacks such molecules and bring ionization. In this paper, numbers of substituted nitrogen atom will be increased from one to four, and ionized from null to tri-cation. Theoretical IR were compared with astronomically well observed one.

Here, we like to show a simple capable story of chemical evolution step of PAH's in interstellar space. One or two nitrogen substituted carbon pentagon-hexagon molecules may have a potential to be created in interstellar space, whereas more than three nitrogen atoms substituted one may be not.

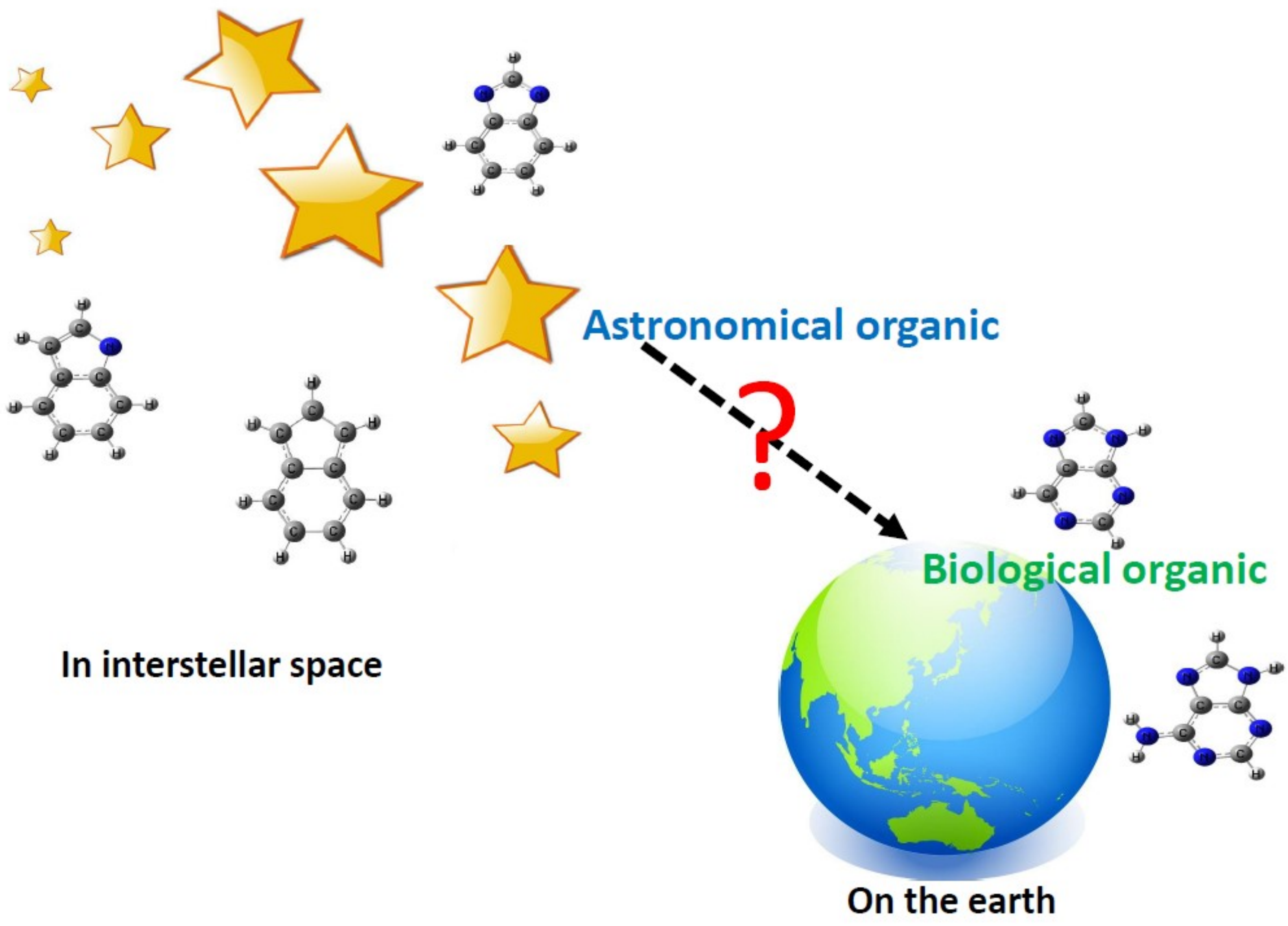

Figure 1, Image of creation of Polycyclic Aromatic Hydrocarbon (PAH) in interstellar space as "astronomical organic. Carbon pentagon-hexagon molecules are good candidates. Whereas on the earth, there are many biological elements with pentagon-hexagon structure as like purine and adenine. Capability of chemical evolution step between those two groups was not yet known.

## 2, MODELING NITROGEN SUBSTITUTED PAH

One typical example of model molecule was illustrated in Figure 2. Starting molecule is carbon pentagon-hexagon molecule with $C_9H_7$. In interstellar space, especially in new star born area, high energy nitrogen and high energy photon may attack such simple molecule. Particular carbon site of $C_9H_7$ will be substituted by nitrogen as like $C_8H_6N_1$. Such modified molecule may be ionized by high energy photon to be cation as $C_8H_6N_1^{+m}$. By presuming such simple process, we can illustrate nitrogen substitution step and model molecules as Figure 3. None substituted case is shown as N=0, to be $C_{10}H_8$ (naphthalene) and $C_9H_7$ (Ota, 2015c and 2015d). One nitrogen substituted case of N=1 was divided to two cases as site-a substituted $C_8H_6N_1$-a and site-b $C_8H_6N_1$-b. Whereas, two nitrogen substituted cases are test as $C_7H_5N_2$-bb and $C_7H_5N_2$-ab.

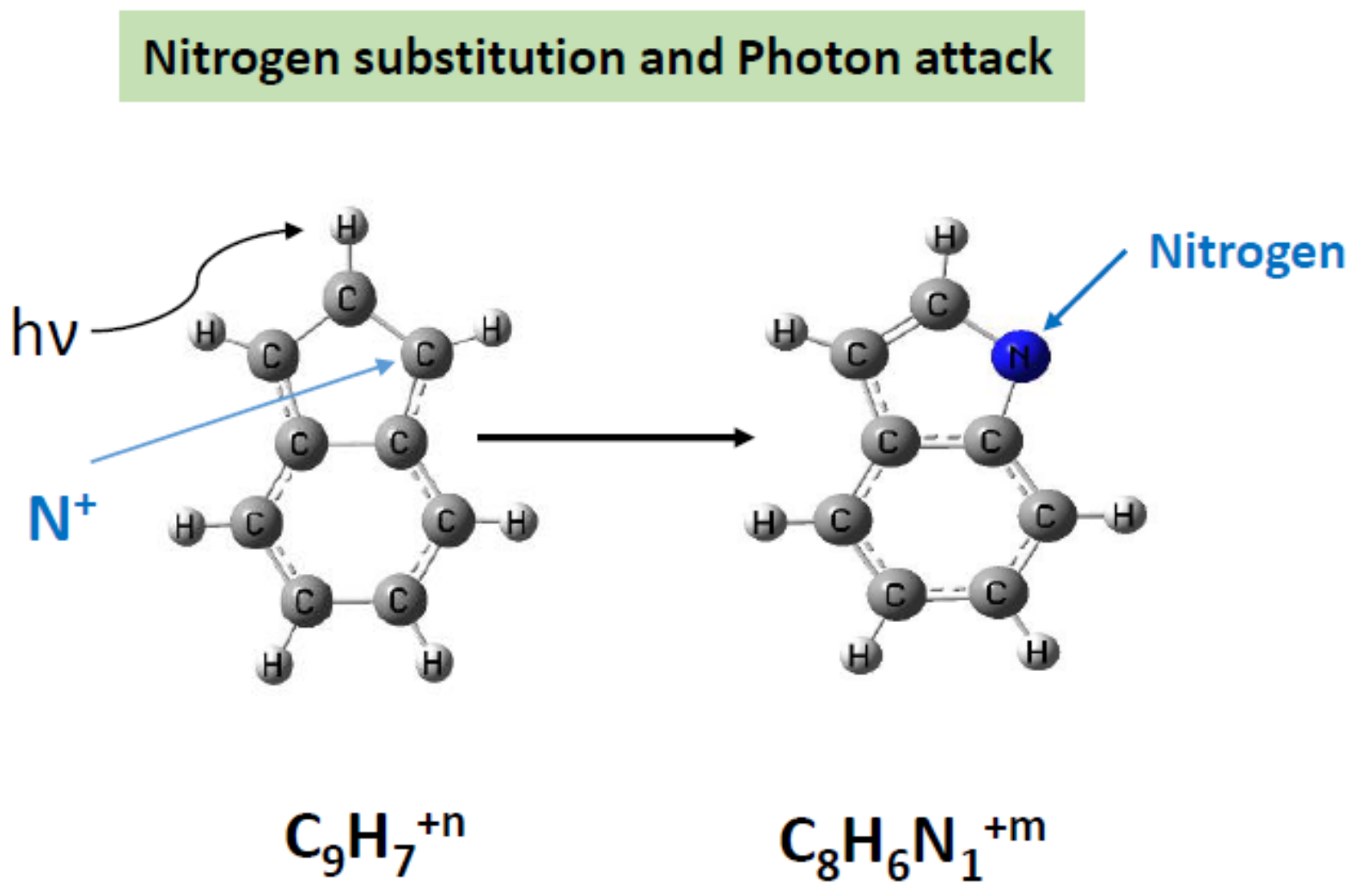

Figure 2, Modeling nitrogen substituted PAH's. Carbon site of pentagon-hexagon molecule $C_9H_7$ will be substituted by high energy nitrogen, and ionized by high energy photon.

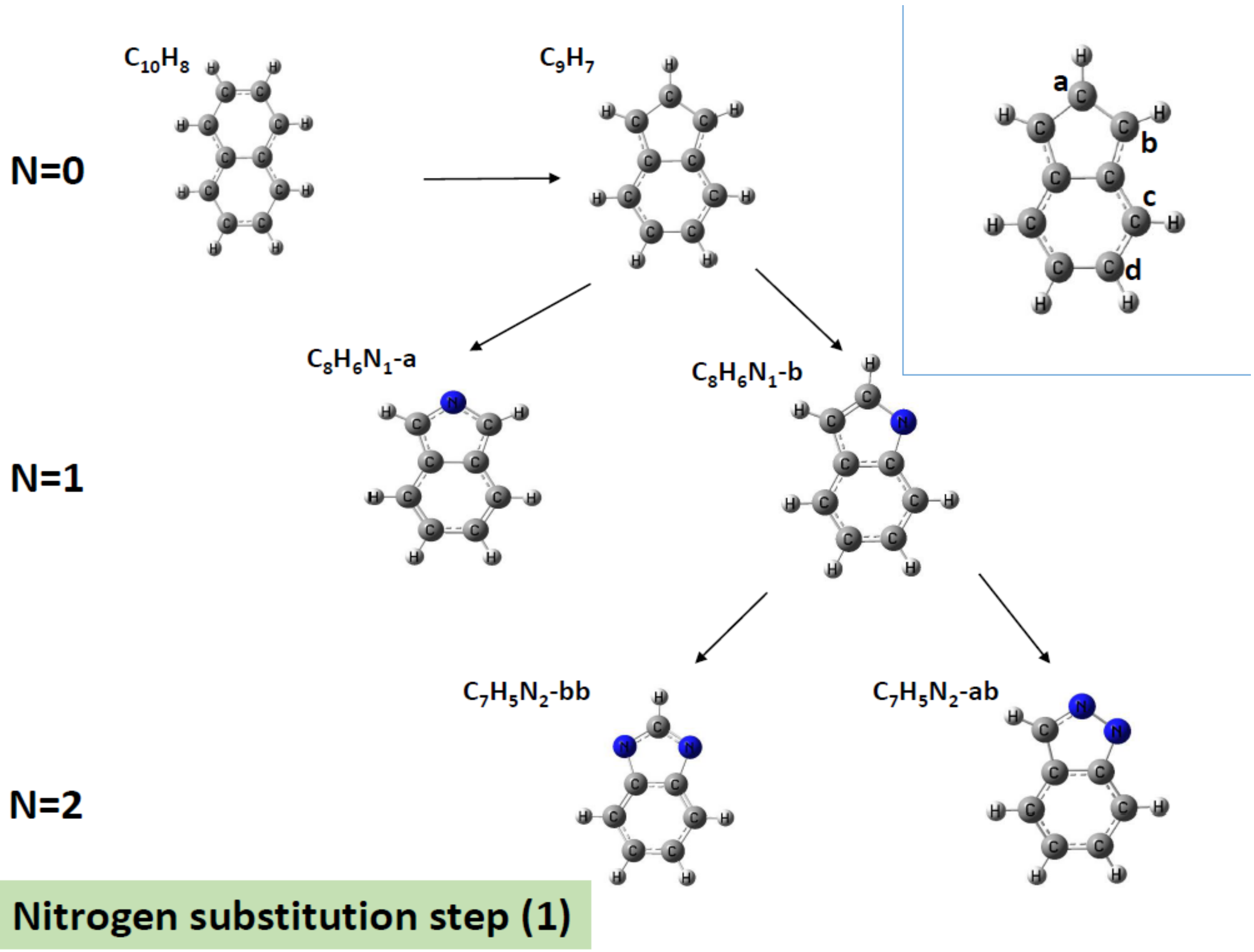

Figure 3, Nitrogen substitution step for carbon pentagon-hexagon molecule group. Blue ball show nitrogen atom. Notation of carbon sites are labelled in right upper corner as a, b, c, and d.

## 3, CALCULATION METHOD

We have to obtain total energy, optimized atom configuration, and infrared vibrational mode frequency and strength depend on a given initial atomic configuration, charge and spin state Sz. Density functional theory (DFT) with unrestricted B3LYP functional (Becke 1993) was applied utilizing Gaussian09 package (Frisch et al. 2009, 1984) employing an atomic orbital 6-31G basis set. The first step calculation is to obtain the self-consistent energy, optimized atomic configuration and spin density. Required convergence on the root mean square density matrix was less than $10^{-8}$ within 128 cycles. Based on such optimized results, harmonic vibrational frequency and strength was calculated. Vibration strength is obtained as molar absorption coefficient ε (km/mol.). Comparing DFT harmonic wavenumber $N_{DFT}$ ($cm^{-1}$) with experimental data, a single scale factor 0.965 was used (Ota 2015b). For the anharmonic correction, a redshift of 15$cm^{-1}$ was applied (Ricca et al. 2012).

Corrected wave number N is obtained simply by $N\ (cm^{-1}) = N_{DFT}\ (cm^{-1})\ x\ 0.965 - 15\ (cm^{-1})$.

Also, wavelength λ is obtained by $\lambda\ (\mu m) = 10000/N(cm^{-1})$.

## 4, ONE AND TWO NITROGEN SUBSTITUTED MOLECULES

### 4.1 $C_8H_6N_1$-a

Calculated IR spectrum of site-a substituted molecule $C_8H_6N_1$-a was illustrated in Figure 4 varying charge number from 0 to +3. Green broken lines were astronomically well observed wavelength as 3.3, 6.2, 7.6, 7.8, 8.6, 11.2, 12.7, 13.5, and 14.3 micrometer. It is obvious that in every charged case calculated spectrum show no good resemblance with observed lines, that is, major observed peaks of 6.2, 7.6, 7.8 and 11.2 micrometer could not be reproduced.

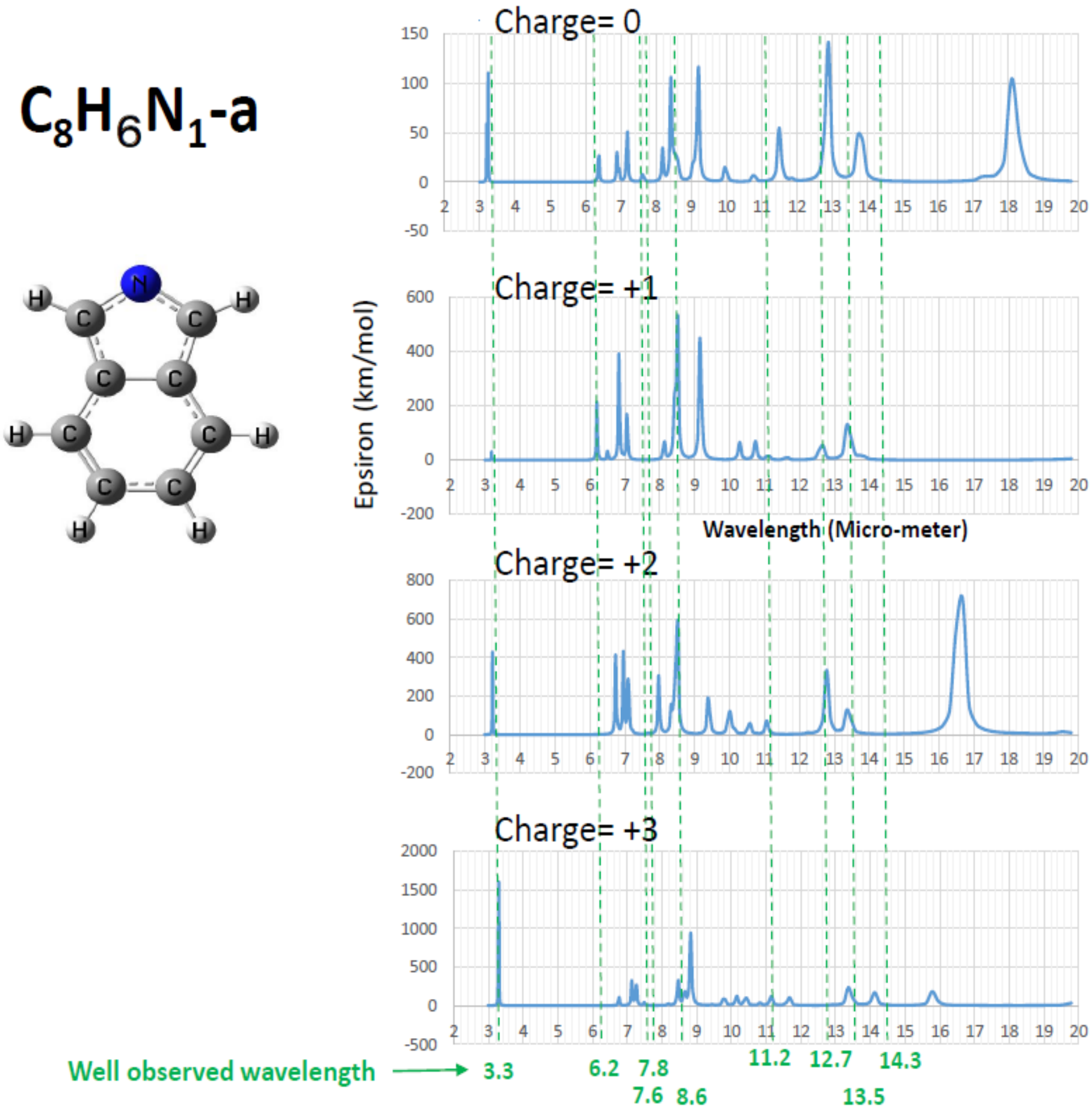

Figure 4, Calculated IR spectra of $C_8H_6N_1$-a drawn by blue curve. We could not see any good coincidence with astronomically well observed peaks marked by green broken lines.

4.2 **$C_8H_6N_1$-b**

Site-b substituted molecule $C_8H_6N_1$-b show a remarkable spectra as shown in Figure 5. Especially, in case of charge= +1 (mono cation), major observed lines of 3.3, 6.2, 7.6, and 8.6 micrometer were well reproduced within a range of +/- 0.2 micrometer. Fundamental vibration mode analysis was performed as summarized in Table 1, where left side green numbers were well observed wavelength. Mode number, wavelength, and epsiron (vibrational strength) were tabled. In column of epsiron, blue numbers show good coincidence with observed mode, whereas red one no good. We can judge that site-b substituted mono cation $C_8H_6N_1$-$b^{+1}$ is one candidate to have a capability existing in interstellar space.

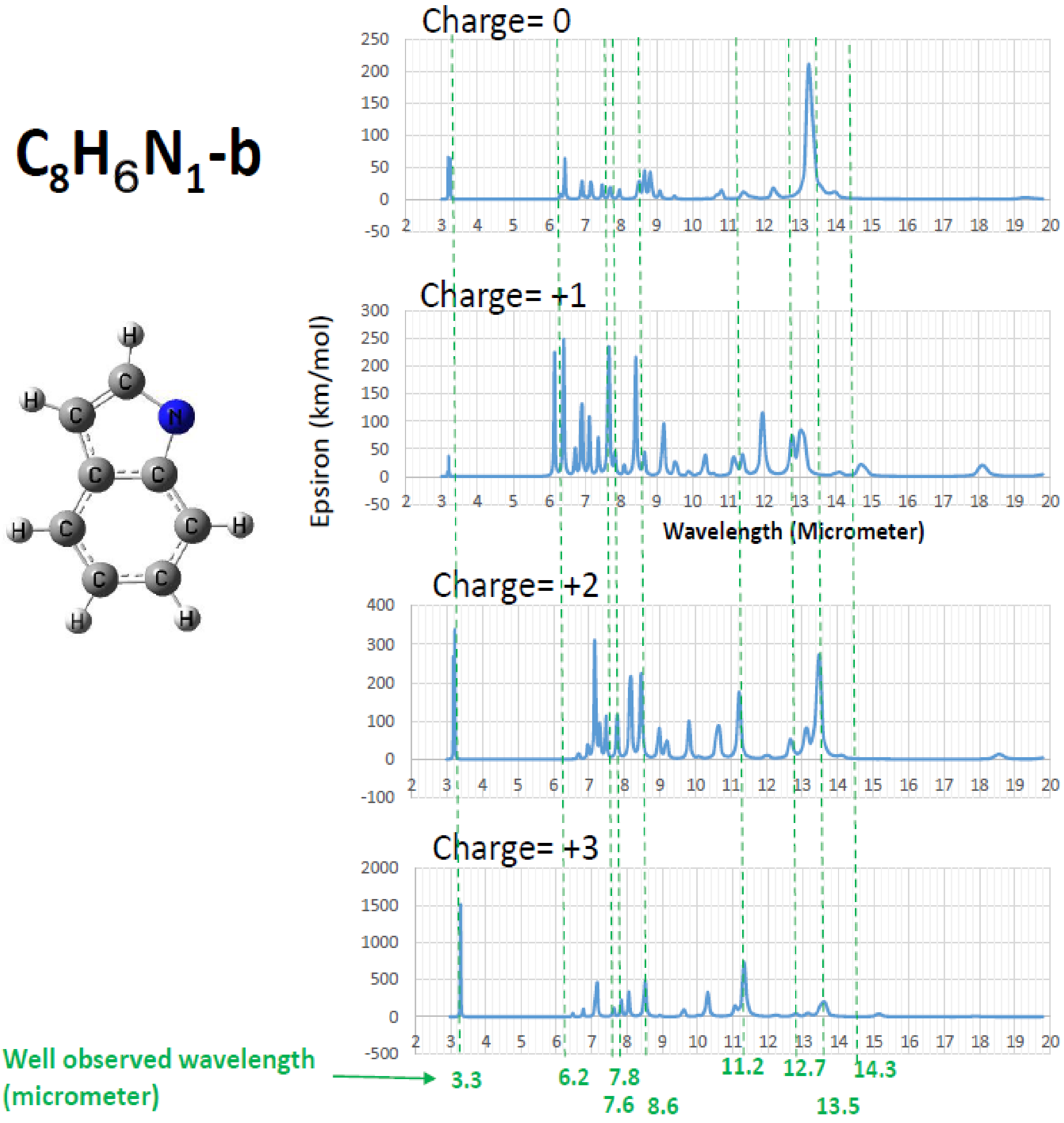

Figure 5, IR spectra of site-b substituted molecule $C_8H_6N_1$-b. In case of charge=+1, we can see good coincidence of calculated spectrum (blue curve) with observed one (green broken lines).

Table 1, Mode analysis of $C_8H_6N_1$-b. Good coincident modes were marked by blue number in column of IR intensity epsiron. Whereas, no good coincident or calculated modes were marked by red. Major IR behavior was reproduced fairly good.

$C_8H_6N_1$-b
Charge=+1

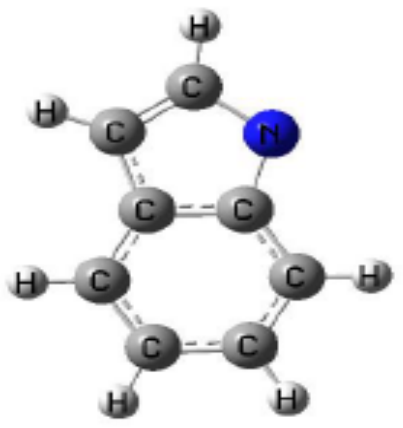

| Observed wavelength (micron) | Calculated IR spectrum | | | |
|---|---|---|---|---|
| | Mode number | Wavelength (micron) | Epsiron (km/mol) | Vibrational mode |
| | 1 | 70.8 | 15.0 | |
| | 2 | 48.2 | 2.3 | |
| | 3 | 28.5 | 4.8 | |
| | 4 | 26.6 | 5.3 | |
| | 5 | 22.5 | 4.0 | |
| | 6 | 20.9 | 4.0 | |
| | 7 | 19.9 | 1.7 | |
| | 8 | 18.1 | 6.1 | |
| | 9 | 14.7 | 9.1 | C-C stretching, C-H in plane bending |
| 14.3 | 10 | 14.1 | 2.7 | |
| 13.5 | 11 | 13.1 | 37.8 | C-H out of plane bending |
| 12.7 | 12 | 12.8 | 19.1 | C-C stretching, C-H in plane bending |
| | 13 | 12.0 | 3.6 | |
| | 14 | 12.0 | 31.3 | C-H out of plane bending |
| | 15 | 11.4 | 10.6 | |
| 11.2 | 16 | 11.2 | 12.6 | C-H out of plane bending |
| | 17 | 10.6 | 1.1 | |
| | 18 | 10.3 | 13.3 | |
| | 19 | 10.2 | 0.9 | |
| | 20 | 9.9 | 3.3 | |
| | 21 | 9.5 | 11.1 | |
| | 22 | 9.2 | 34.3 | C-H in plane bending |
| | 23 | 8.7 | 14.4 | |
| 8.6 | 24 | 8.4 | 63.3 | C-H in plane bending |
| | 25 | 8.1 | 5.4 | |
| 7.8 | 26 | 7.8 | 11.1 | |
| 7.6 | 27 | 7.7 | 100.6 | C-H in plane bending |
| | 28 | 7.4 | 23.0 | |
| | 29 | 7.1 | 30.9 | |
| | 30 | 6.9 | 60.4 | C-H in plane bending |
| | 31 | 6.7 | 18.7 | |
| | 32 | 6.4 | 109.8 | C-C stretching, C-H in plane bending |
| 6.2 | 33 | 6.2 | 67.3 | C-C stretching, C-H in plane bending |
| | 34 | 3.2 | 0.3 | |
| | 35 | 3.2 | 1.6 | |
| | 36 | 3.2 | 5.7 | |
| | 37 | 3.2 | 0.0 | |
| 3.3 | 38 | 3.2 | 8.6 | C-H stretching |
| | 39 | 3.2 | 3.5 | |

### 4.3 $C_7H_5N_2$-ab

Two nitrogen atoms were substituted to original $C_9H_7$ molecule to be $C_7H_5N_2$-ab, of which calculated IR were compared in Figure 6 varying charge number. From charge 0 to +2, there were no good IR behavior resemble to observed one, whereas good coincidence in case of charge= +3. Bonding between two nitrogen atoms suggested to bring such difference, that is, the former bond was single bond, the latter double bond. In case of charge= +3, two nitrogen are bonded tightly and act as one heavy center of vibration. For molecular vibration, it looks similar atomic configuration with $C_8H_6N_1$-b having charge=+1.

Vibration mode analysis was summarized in Table 2. Five modes with strong C-H stretching modes were obtained around 3.3 micrometer. Also, strong vibration at 6.4, 7.6, 8.5, 11.4, and 14.3 micrometer, which were correspond to observed 3.3, 6.2, 7.6, 8.6, 11.2, and 14.3 micrometer. However, calculated 7.3 and 9.1 micrometer mode were not known by common observation.

### 4.4 $C_7H_5N_2$-bb

Two site-b carbons were substituted by two nitrogen as $C_7H_5N_2$-bb. Calculated result of charge=+1 gave a good example as shown in Figure 7. Detailed vibration analysis was noted in Table 2, where well observed 3.3, 6.2, 7.6, 8.6, 11.2, and 12.7 micrometer wavelength were reproduced well within +/- 0.2 micrometer as marked by blue number. There were some extra wavelength of 6.8, 7.2 and 11.7 micrometer not observed in interstellar space. There was no mode at 13.5 micrometer in calculation.

### 4.5 Summary of N=0, 1, and 2 cases

In Table 4, calculated IR's were summarized. Substituted number of nitrogen atoms were N=0, 1 and 2. In each molecule box, charge numbers were inserted, where black number case means no good coincidence between observed IR and calculated one, also green number show fair coincidence, and blue number good.

**4.6 Comparison of IR spectra with interstellar observation**

Beautiful typical interstellar IR observation results were edited by NASA group (Boersma et al, 2009) using four astronomical sources as shown on a top of Figure 8, Calculated three good candidates are illustrated as $C_8H_6N_1$-b$^{+1}$, $C_7H_5N_2$-bb$^{+1}$, and $C_7H_5N_2$-ab$^e$. Among those three, the best one will be $C_7H_5N_2$-bb$^{+1}$, which wavelength and relative IR strength could almost reproduce observed feature from 3 to 11 micrometer. Unfortunately at longer wavelength than 11 micrometer, three calculated examples could not reproduce well by such nitrogen substituted two carbon ring molecule group. We should study other advanced system including three, four and more carbon rings as like discussed in previous papers (Ota 2014, 2015d).

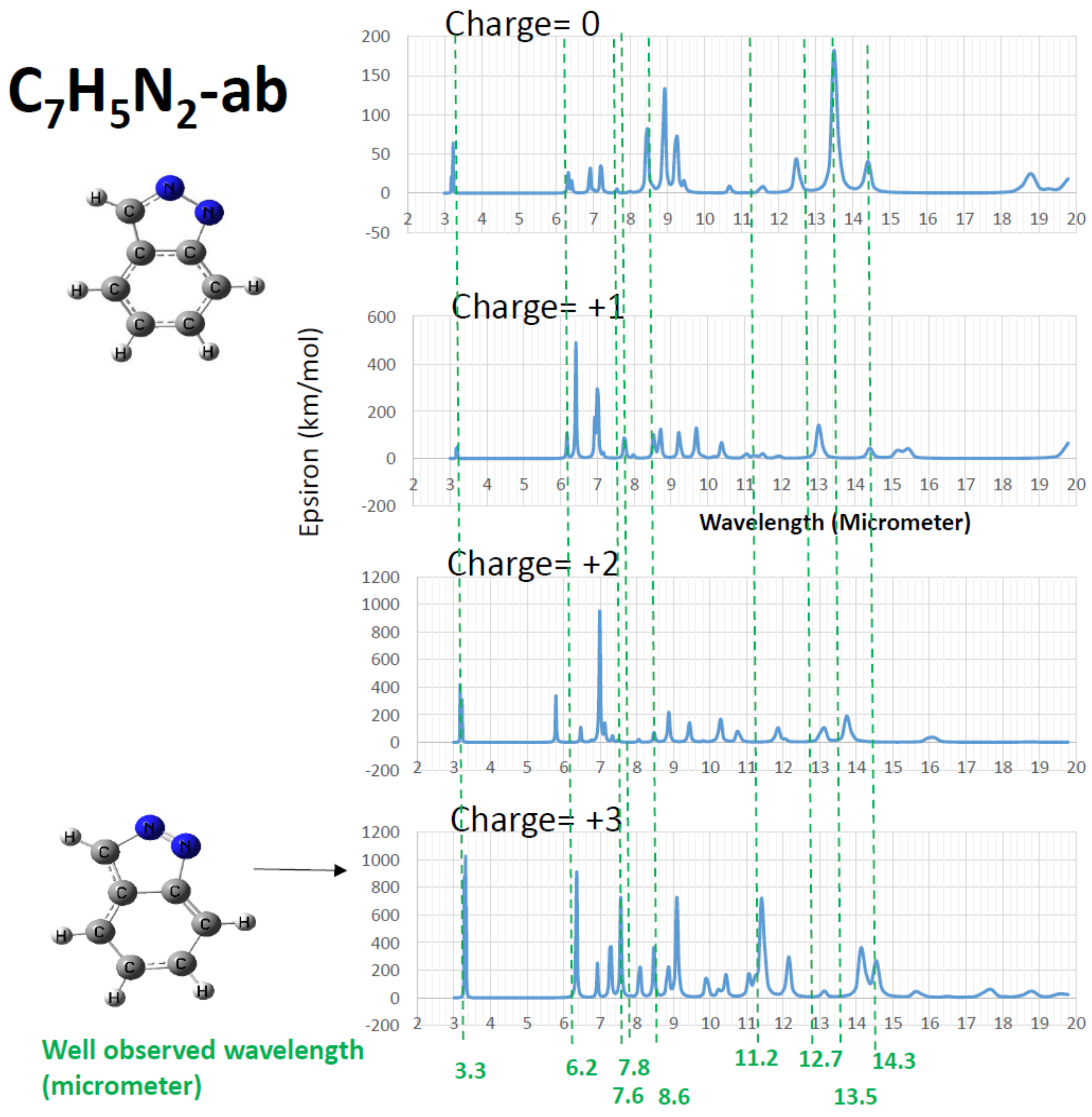

Figure 6, Calculated IR spectra of $C_7H_5N_2$-ab. In case of charge=+3, observed IR was reproduced fairly nice.

Table 2, Mode analysis of $C_7H_5N_2$-ab$^{3+}$. Good coincident modes were marked by blue number, of which major IR behavior was reproduced fairly nice. Red marked mode were not known compared with well observed wavelength marked by green wavelength.

**$C_7H_5N_2$-ab**

**Charge=+3**

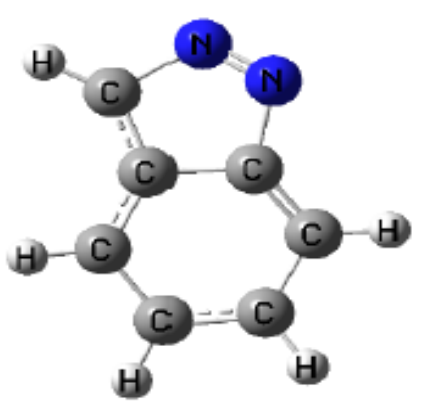

| Observed wavelength (micro meter) | Calculated IR Spectrum | | | |
|---|---|---|---|---|
| | Mode number | Wavelength (micro meter) | IR intensity (km/mol) | Vibrational mode |
| | 1 | 64.4 | 21.0 | |
| | 2 | 58.8 | 2.3 | |
| | 3 | 49.7 | 18.5 | |
| | 4 | 34.2 | 5.4 | |
| | 5 | 27.6 | 25.3 | |
| | 6 | 25.1 | 33.8 | |
| | 7 | 19.7 | 12.1 | |
| | 8 | 18.7 | 15.2 | |
| | 9 | 17.6 | 23.3 | |
| | 10 | 16.5 | 2.2 | |
| | 11 | 15.7 | 16.3 | |
| | 12 | 14.5 | 72.1 | |
| 14.3 | 13 | 14.2 | 122.9 | C-H out of plane bending |
| 12.7, 13.5 | 14 | 13.1 | 13.6 | |
| | 15 | 12.1 | 85.6 | |
| 11.2 | 16 | 11.4 | 260.6 | C-C, C-N stretching, C-H out of plane bending |
| | 17 | 11.3 | 31.6 | |
| | 18 | 11.1 | 44.9 | |
| | 19 | 10.4 | 51.0 | |
| | 20 | 10.2 | 19.0 | |
| | 21 | 9.9 | 60.1 | |
| | 22 | 9.1 | 218.4 | C-H in-plane bending at pentagon |
| | 23 | 8.9 | 87.8 | |
| 8.6 | 24 | 8.5 | 104.3 | C-H in-plane bending |
| | 25 | 8.1 | 93.8 | |
| 7.8 | 26 | 7.8 | 2.0 | |
| 7.6 | 27 | 7.6 | 214.8 | C-H in-plane bending |
| | 28 | 7.3 | 135.9 | C-H in-plane bending |
| | 29 | 7.3 | 44.4 | |
| | 30 | 6.9 | 72.5 | |
| 6.2 | 31 | 6.4 | 339.2 | C-C, C-N stretching |
| | 32 | 3.3 | 138.3 | C-H stretching at hexagon |
| | 33 | 3.3 | 118.5 | C-H stretching at hexago |
| 3.3 | 34 | 3.3 | 293.0 | C-H stretching at hexagon |
| | 35 | 3.3 | 226.0 | C-H stretching at hexagon |
| | 36 | 3.3 | 247.3 | C-H stretching at pentagon |

**$C_7H_5N_2$-bb**

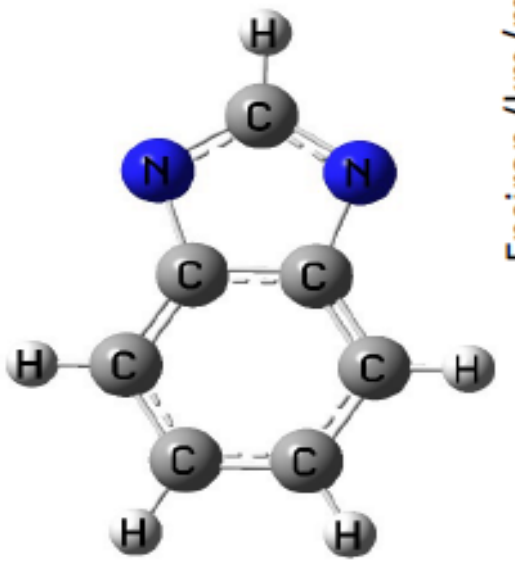

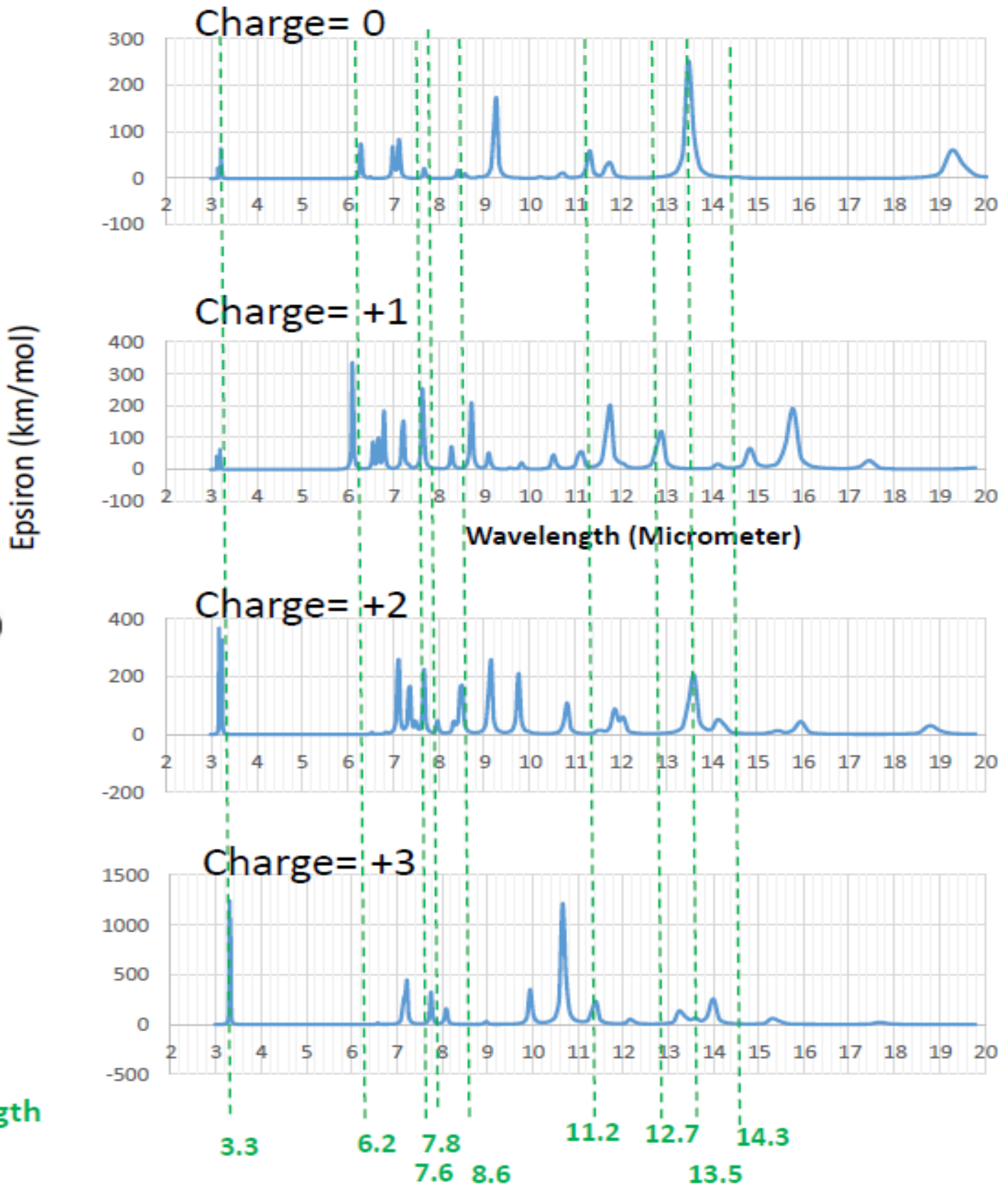

Well observed wavelength (micrometer)

Figure 7, Calculated IR spectra of $C_7H_5N_2$-bb. In case of charge=+1, calculated IR was reproduced fairly nice.

Table 3, Mode analysis of $C_7H_5N_2$-bb$^+$. Good coincident mode was marked by blue, of which major IR behavior was reproduced fairly nice. Not coincident mode were marked by red.

**$C_7H_5N_2$-bb**
**Charge=+1**

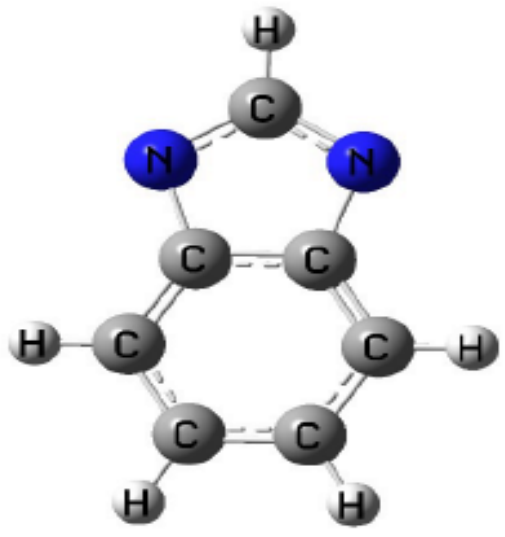

| Observed Wavelength | Calculated IR Spectrum | | | |
|---|---|---|---|---|
| | Mode number | Wavelength (micro meter) | IR intensity (km/mol) | Vibrational mode |
| | 1 | 79.9 | 1.5 | |
| | 2 | 49.0 | 0.4 | |
| | 3 | 29.8 | 0.0 | |
| | 4 | 27.0 | 17.7 | |
| | 5 | 24.2 | 25.1 | C-C out of plane bending |
| | 6 | 20.4 | 7.6 | |
| | 7 | 20.3 | 13.3 | |
| | 8 | 17.4 | 8.2 | |
| | 9 | 15.7 | 64.2 | C-C stretching at pentagon |
| | 10 | 14.8 | 18.8 | |
| 14.3 | 11 | 14.1 | 4.6 | C-C stretching |
| 13.5 | | | none | |
| 12.7 | 12 | 12.9 | 46.8 | C-H out of plane bending |
| | 13 | 12.0 | 4.4 | |
| | 14 | 11.7 | 76.9 | C-C stretching |
| | 15 | 11.6 | 1.9 | |
| 11.3 | 16 | 11.1 | 24.1 | C-H out of plane bending |
| | 17 | 10.5 | 14.8 | |
| | 18 | 10.1 | 0.1 | |
| | 19 | 9.8 | 6.6 | |
| | 20 | 9.6 | 1.5 | |
| | 21 | 9.1 | 16.5 | |
| 8.6 | 22 | 8.7 | 60.8 | C-C stretching |
| | 23 | 8.3 | 20.6 | |
| | 24 | 8.0 | 0.4 | |
| 7.6, 7.8 | 25 | 7.7 | 89.8 | C-C stretching |
| | 26 | 7.3 | 3.3 | |
| | 27 | 7.2 | 68.2 | C-C stretching |
| | 28 | 6.8 | 54.1 | C-C stretching |
| | 29 | 6.7 | 29.5 | |
| | 30 | 6.6 | 26.0 | |
| 6.2 | 31 | 6.1 | 133.1 | C-C stretching |
| | 32 | 3.2 | 1.7 | |
| | 33 | 3.2 | 8.9 | |
| 3.3 | 34 | 3.2 | 14.3 | C-H stretching at hexagon |
| | 35 | 3.2 | 5.7 | |
| | 36 | 3.1 | 13.7 | C-H stretching at pentagon |

Table 4, Summary of calculated IR compared with astronomically well observed one. Substituted number of nitrogen atoms were N=0, 1 and 2. Black number charged case show no good coincidence, green number fair, blue number good.

| | | | |
|---|---|---|---|
| **N=0** | $C_{10}H_8$ 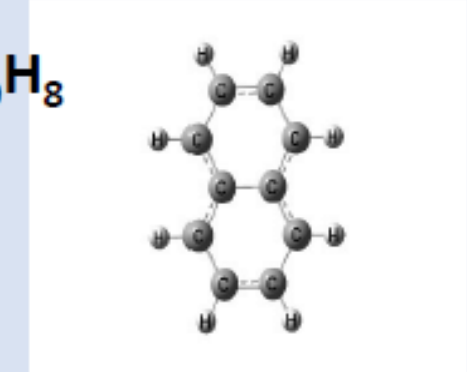 Charge: 0, +1, +2, +3 | $C_9H_9$ 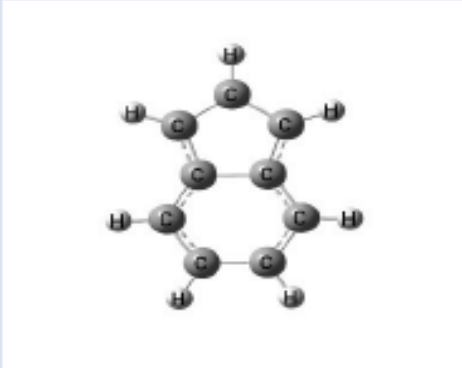 0, +1, +2, +3 | |
| **N=1** | | $C_8H_6N_1$-a 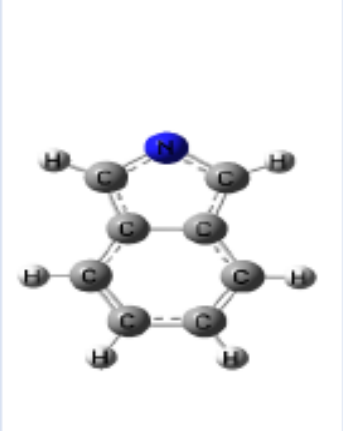 Charge: 0, +1, +2, +3 | $C_8H_6N_1$-b 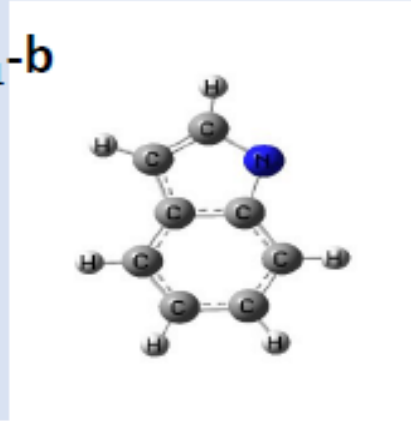 0, +1, +2, +3 |
| **N=2** | | $C_7H_5N_2$-bb 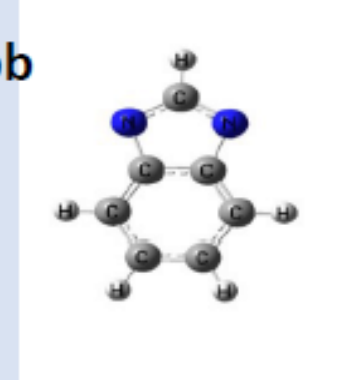 0, +1, +2, +3 | $C_7H_5N_2$-ab 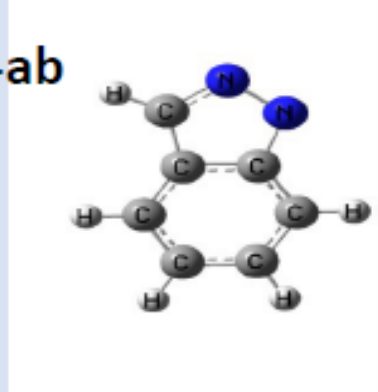 0, +1, +2, +3 |

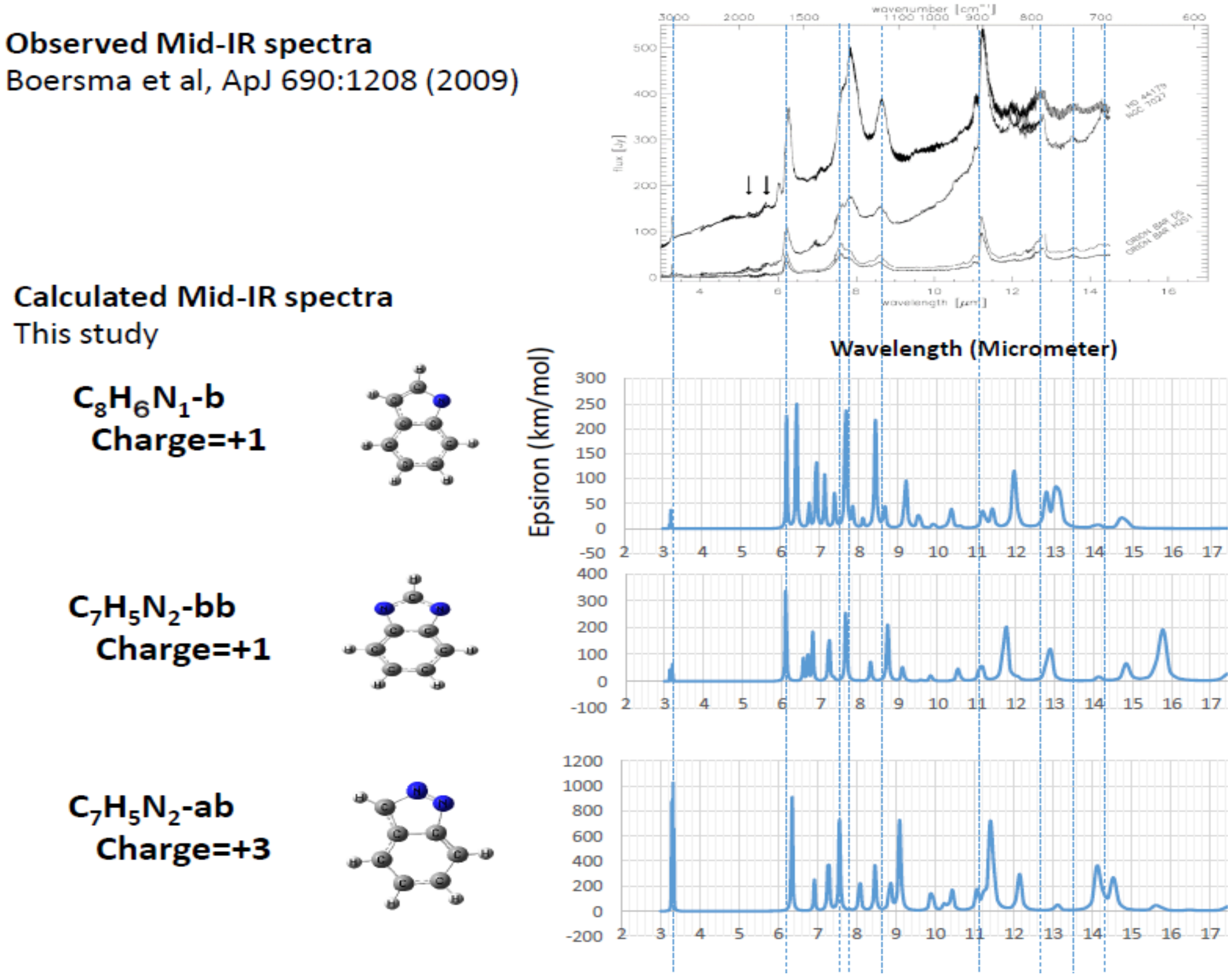

Figure 8, Comparison of IR spectra with observed interstellar four sources (Boersma et al, 2009) and calculated three examples showing good resemblance.

## 5, IR SPECTRA OF THREE AND FOUR NITROGEN SUBSTITUTED MOLECULES

Typical biological molecules having carbon pentagon-hexagon skeleton are purine and adenine, which include four and five nitrogen atoms in its structure. Question is a capability to show similar behavior with astronomically observed IR, in other words, capability of creation in interstellar space. We should study three and four nitrogen substituted candidates as illustrated in Figure 9. Starting molecule was $C_7H_5N_2$-bb, of which c or d site will be substituted to be $C_6H_4N_3$-bbc or $C_6H_4N_3$-bbd. In case of four nitrogen, there will be checked three candidates as $C_5H_3N_4$-bbcc, $C_5H_3N_4$-bbcd, and $C_5H_3N_4$-bbdd.

### 5.1 $C_6H_4N_3$-bbc

Calculated IR of $C_6H_4N_3$-bbc changing charge number from 0 to +3 was shown in Figure 10. In case of charge zero, there was no good coincidence with observed IR except at 3.2 and 7.5 micrometer. Serious molecule configuration change was observed. As shown in the tilt angle view of charge=+1 and +2, molecule was not placed on one plane. In case of charge=+3, two rings were untied to be single ring. In every cases, calculated IR were not coincident with observed one.

### 5.2 $C_6H_4N_3$-bbd

As shown in Figure 11, $C_6H_4N_3$-bbd show no good IR spectra for cases of charge=0, +2 and +3. DFT calculation were not converged in case of charge= +1 due to very serious and unstable configuration change.

### 5.3 $C_5H_3N_4$-bbcc

Additional two nitrogen atoms were substituted to site c to be $C_5H_3N_4$-bbcc as illustrated in Figure 12. Except a case of charge=0, molecular configuration seriously changed to one ring. In any case, there are no good similarity with calculated IR behavior compared with well observed green broken lines.

### 5.4 $C_5H_3N_4$-bbdd

Result of two d sites substituted $C_5H_3N_4$-bbdd was shown in Figure 13. Molecular configuration was also drastically modified. In cases of charge=0 and +3, there were no terminated DFT calculation. Survived case of charge=+1 and +2, IR behavior did not look like observed one.

### 5.5 $C_5H_3N_4$-bbcd

Two nitrogen atoms were substituted to site c and d respectively as $C_5H_3N_4$-bbcd as like Figure 14. We could not obtain terminated calculation in cases of charge=+1 and +2. IR of charge=0 and +3 were completely different with well observed green broken line.

### 5.6 Biological component, purine $C_5H_4N_4$-bb'cd

Typical biological organic purine $C_5H_4N_4$ is a modified molecule based on $C_5H_3N_4$-bbcd, which site b nitrogen is hydrogenated as like $C_5H_4N_4$-bb'cd shown in Figure 9 and Figure 15. In every charged case from 0 to +3, we could get DFT solution, but unfortunately could not have any good IR behavior compared with interstellar observed one. Vibrational mode analysis was summarized in Table 5, which show many extra modes marked by red.

### 5.7 Biological component, adenine $C_5H_5N_5$

Adenine $C_5H_5N_5$ is a key component of DNA, which is $NH_2$ attached purine to site c carbon. Obtained IR in Figure 16 suggested no good coincidence with interstellar observed one. Also, vibrational mode analysis in Table6 suggested again little relationship with astronomically observed spectra.

Results of nitrogen substitution from N=3 and 4 including purine and adenine were summarized in Table 7, which show no good IR to reproduce observed one. Number in every molecule box show charge number, of which black one was no good IR, green number means fairly nice compared with observation. Suffix N means not terminated calculation, D also serious deformation, and M deformed to mono ring from pentagon-hexagon coupled two rings. It should be noted that three and four nitrogen atom substituted molecules could not be a candidate of astronomically created organics.

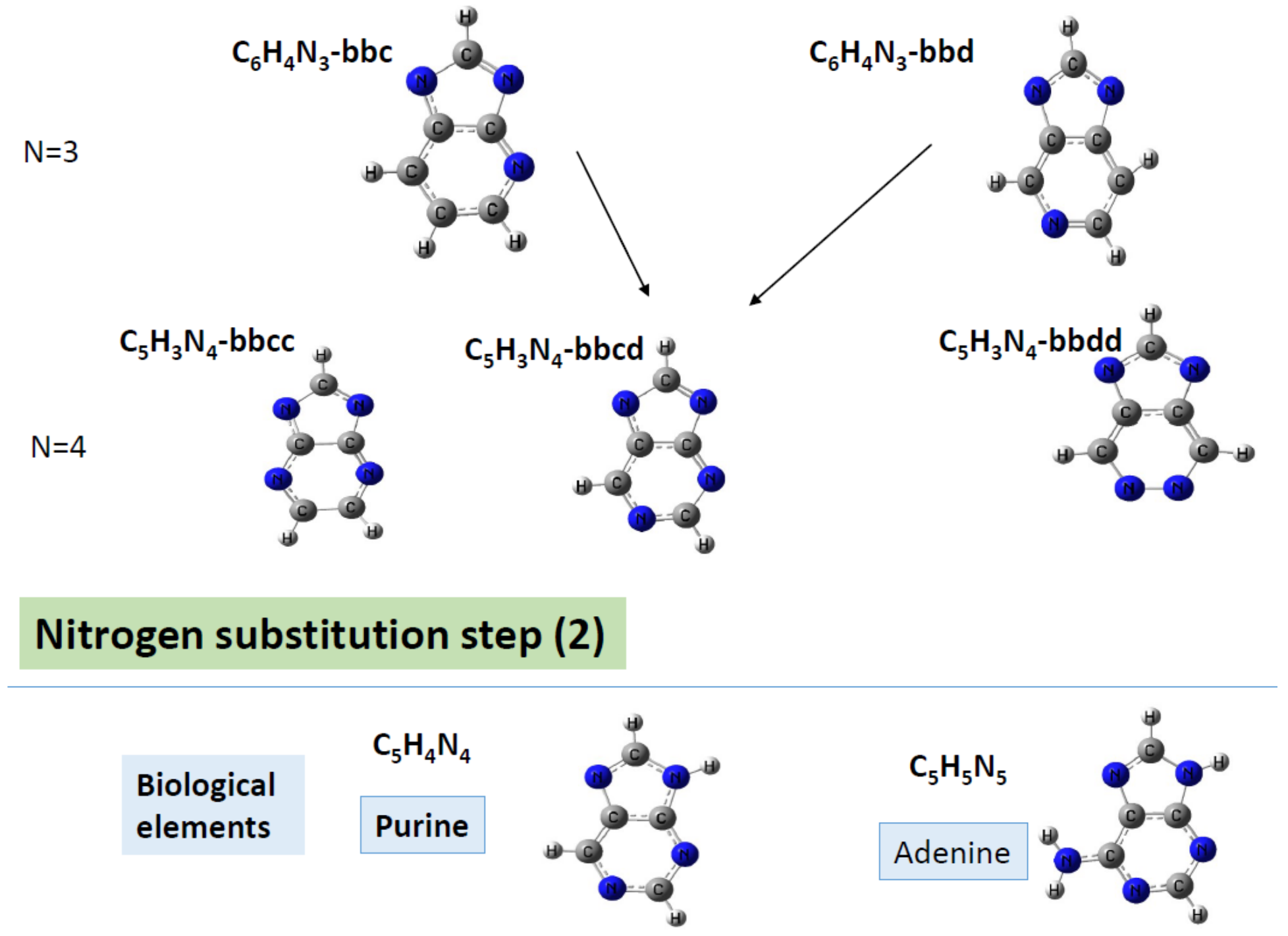

Figure 9, Three and four nitrogen atom substituted pentagon-hexagon coupled molecules including biological organic purine and adenine. Blue ball show nitrogen atom.

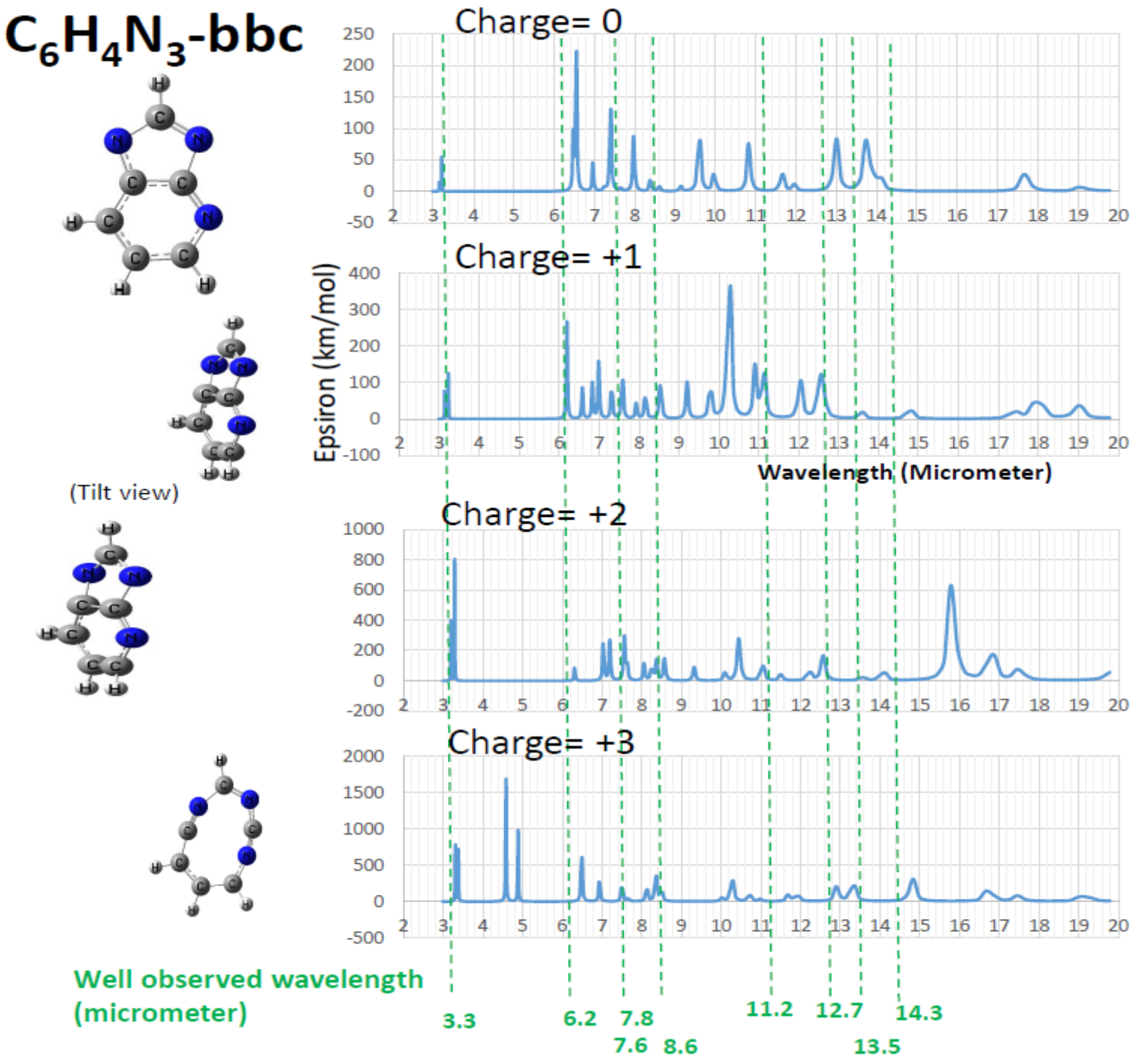

Figure 10, Calculated IR spectra of $C_6H_4N_3$-bbc. Molecular configuration was seriously changed in case of charge=+1, +2, and +3. There are no similar IR behavior with astronomically observed one.

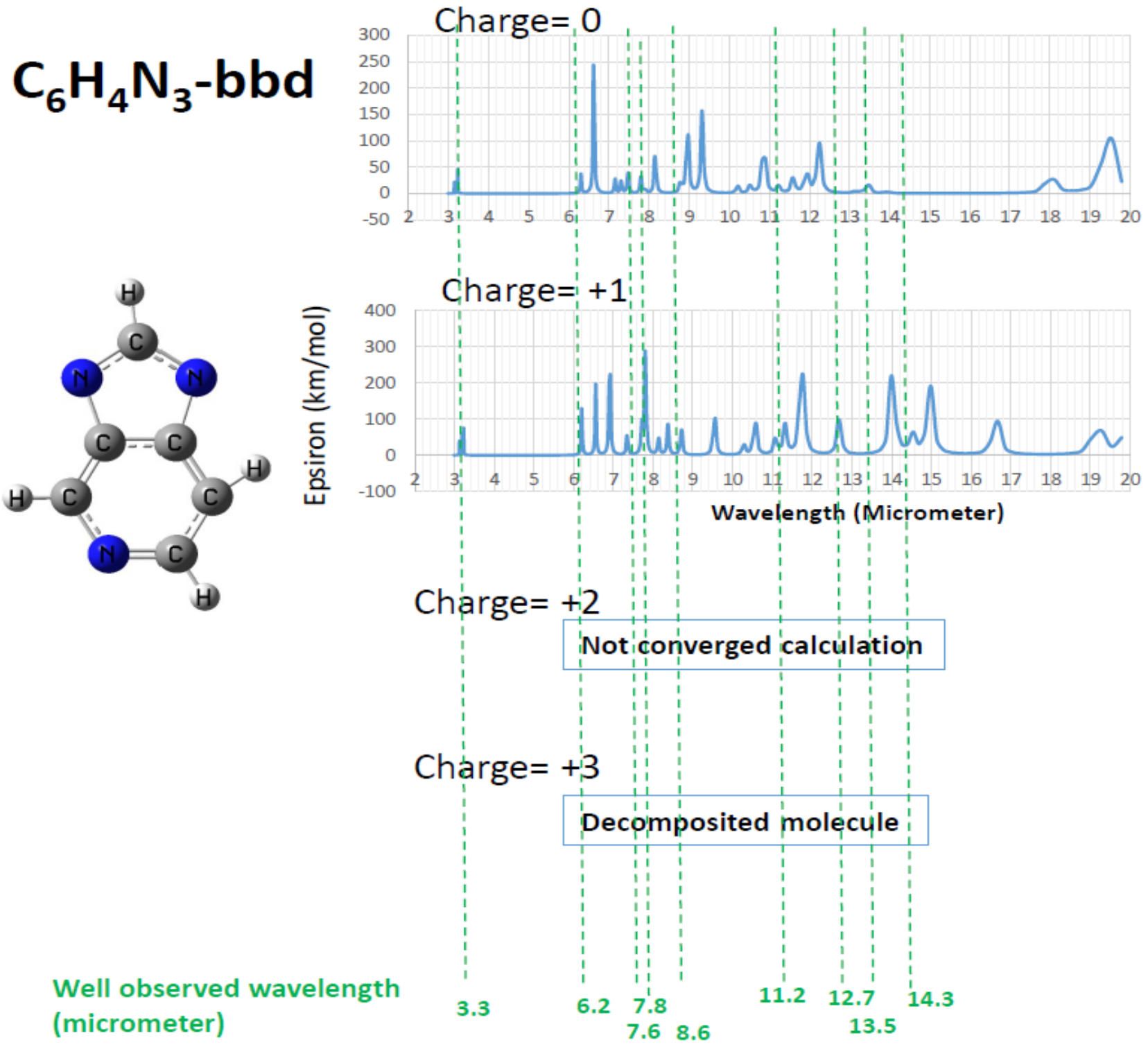

Figure 11, IR spectra of $C_6H_4N_3$-bbd. In cases of charge=0 and +1, calculated spectra gave no good similarity with observed one. DFT calculation were not converged in case of charge=+2 and +3.

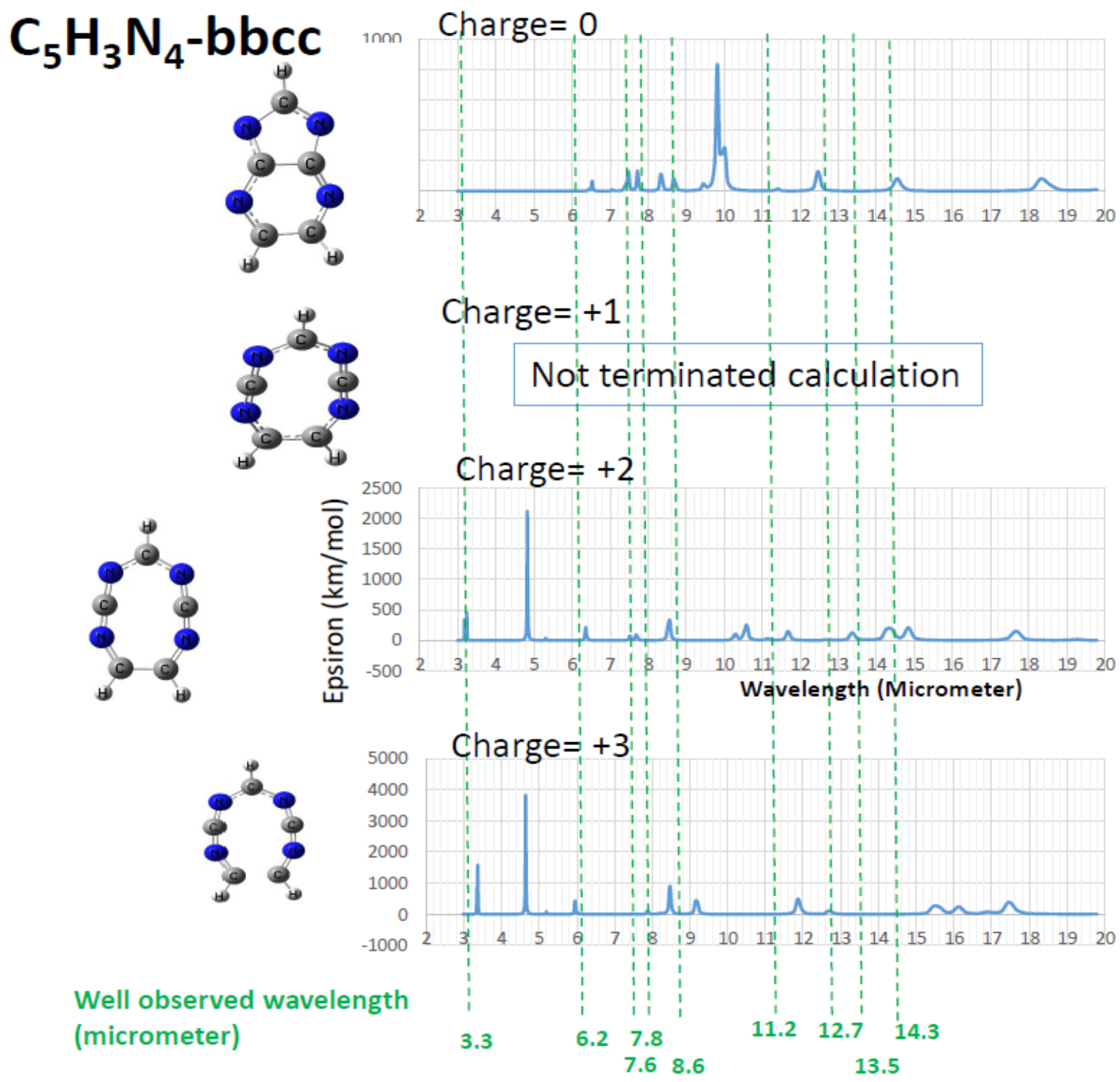

Figure 12, Calculated IR and molecular configuration of $C_5H_3N_4$-bbcc. There are not any good candidates among them.

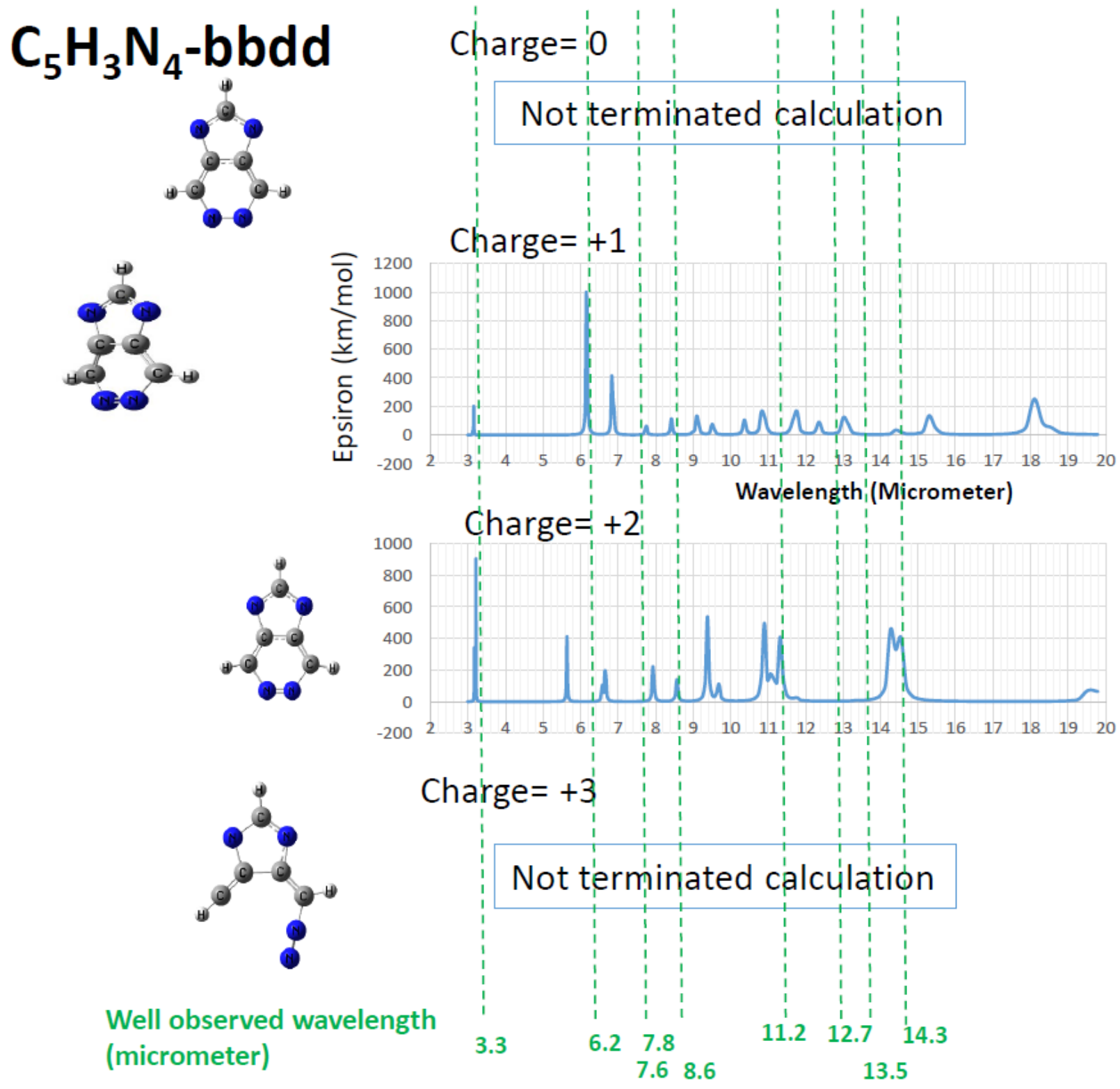

Figure 13, Molecular configuration change and IR of $C_5H_3N_4$-bbdd. IR show no good similarity with observed one.

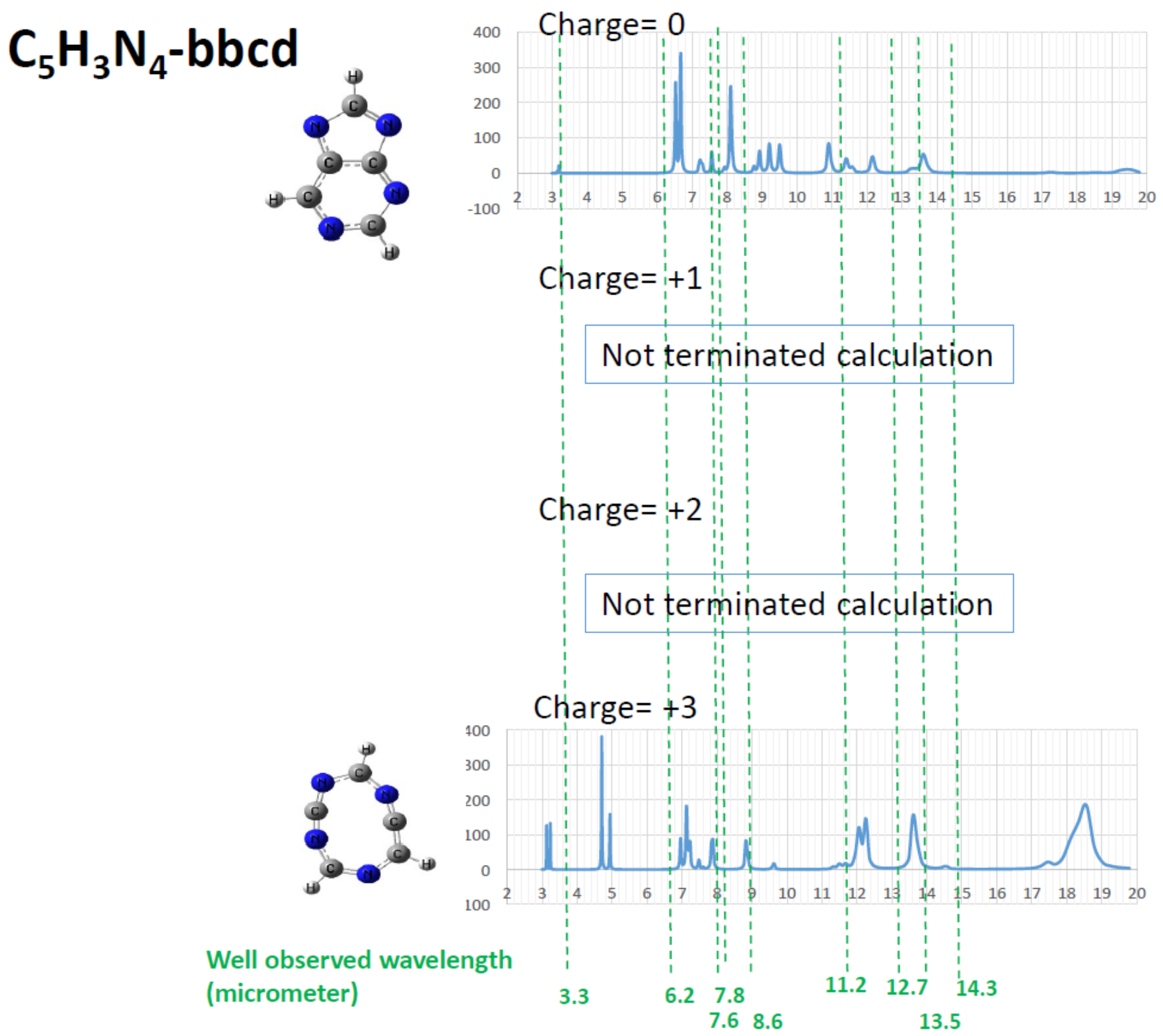

Figure 14, Calculated IR spectra of $C_5H_3N_4$-bbcd. There are not any good similarity with astronomically observed one.

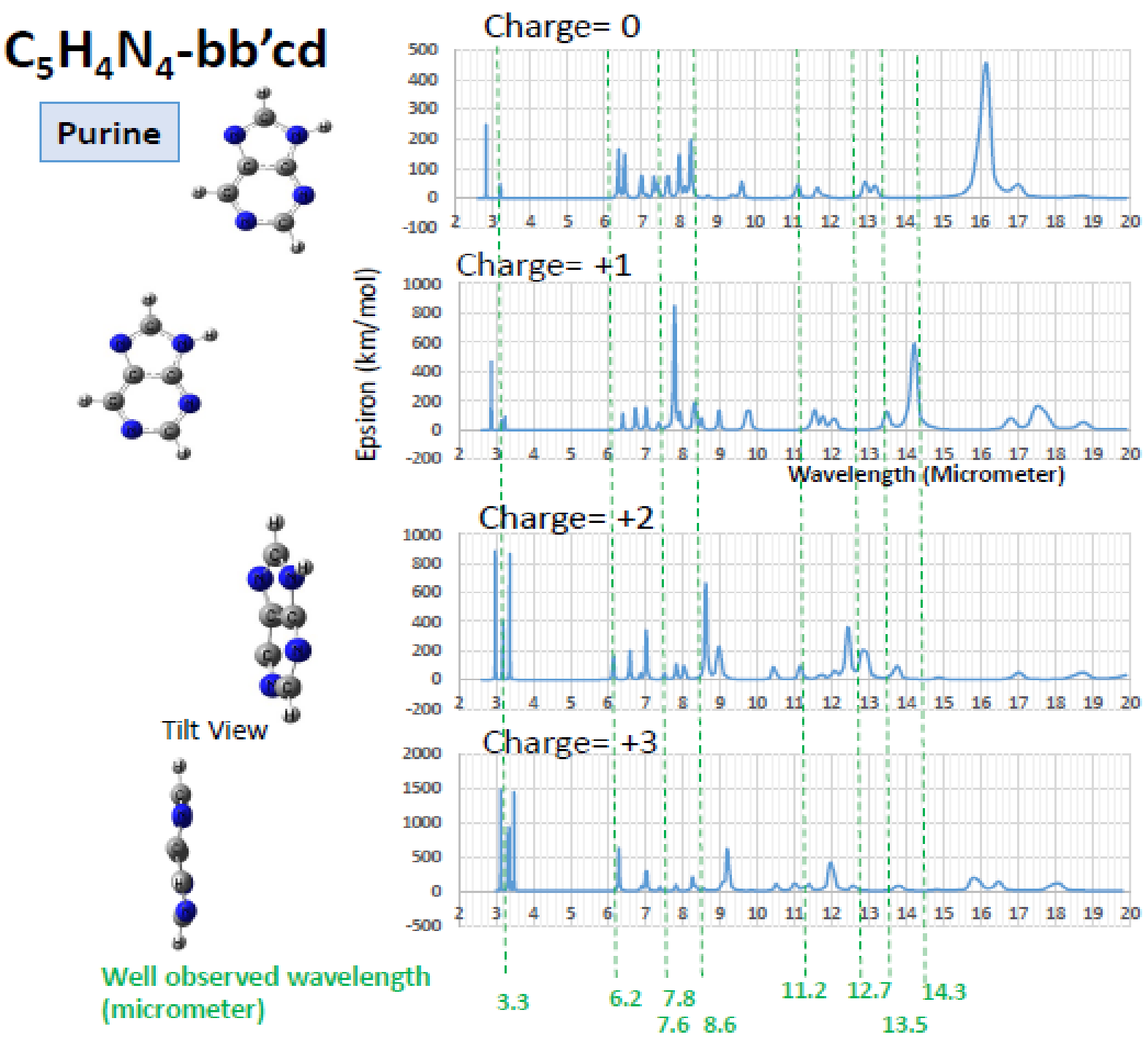

Figure 15, Calculated IR spectra of biological carbon pentagon-hexagon molecule purine as $C_5H_4N_4$-bb'cd. One site-b nitrogen was hydrogenated in $C_5H_3N_4$-bbcd. There are not any good similarity with astronomically observed one marked by green broken line.

Table 5, Vibrational mode analysis of purine with charge=+1. Red marked IR intensity show not coincident mode compared with astronomically well observed wavelength.

$C_5H_4N_4$-bb'cd

Charge=+1

**Purine**

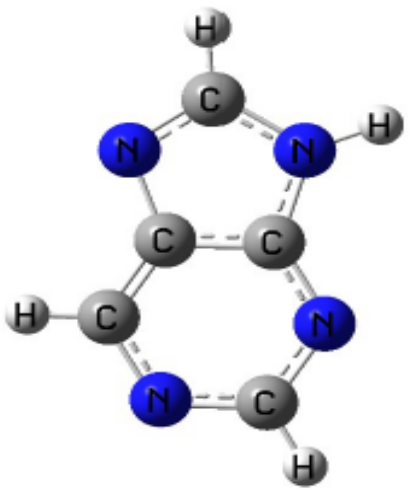

| Observed wavelength (micro meter) | Calculated IR Spectrum | | | |
|---|---|---|---|---|
| | Mode number | Wavelength (micro meter) | IR intensity (km/mol) | Vibrational mode |
| | 1 | 47.9 | 8.7 | |
| | 2 | 43.4 | 10.0 | |
| | 3 | 24.2 | 17.7 | |
| | 4 | 24.2 | 43.1 | |
| | 5 | 21.2 | 124.1 | |
| | 6 | 18.7 | 14.5 | |
| | 7 | 17.6 | 80.1 | |
| | 8 | 16.8 | 23.0 | |
| | 9 | 14.6 | 3.9 | |
| 14.3 | 10 | 14.2 | 193.0 | C–H out of plane bending |
| 13.5 | 11 | 13.5 | 36.2 | C–C stretching |
| 12.7 | | | none | |
| | 12 | 12.1 | 34.7 | |
| | 13 | 11.8 | 24.7 | |
| | 14 | 11.5 | 41.6 | C–C stretching |
| | 15 | 11.5 | 5.0 | |
| 11.2 | 16 | 11.4 | 0.4 | |
| | 17 | 9.8 | 33.8 | C–H in–plane bending, C–N stretching |
| | 18 | 9.7 | 39.7 | C–H in–plane bending, C–N stretching |
| | 19 | 9.0 | 40.1 | C–H in–plane bending, C–N stretching |
| 8.6 | 20 | 8.5 | 22.2 | |
| | 21 | 8.3 | 94.6 | C–H in-plane bending, C–C stretching |
| | 22 | 8.0 | 32.2 | |
| 7.8 | 23 | 7.8 | 269.8 | C–H in-plane bending, C–C stretching |
| 7.6 | 24 | 7.6 | 6.5 | |
| | 25 | 7.4 | 24.5 | |
| | 26 | 7.0 | 50.9 | C–C stretching |
| | 27 | 7.0 | 2.4 | |
| | 28 | 6.8 | 68.0 | C–C stretching |
| 6.2 | 29 | 6.4 | 32.6 | |
| 3.3 | 30 | 3.2 | 48.5 | C–H stretching at hexagon |
| | 31 | 3.1 | 11.1 | |
| | 32 | 3.1 | 12.7 | |
| | 33 | 2.9 | 196.1 | N–H stretching |

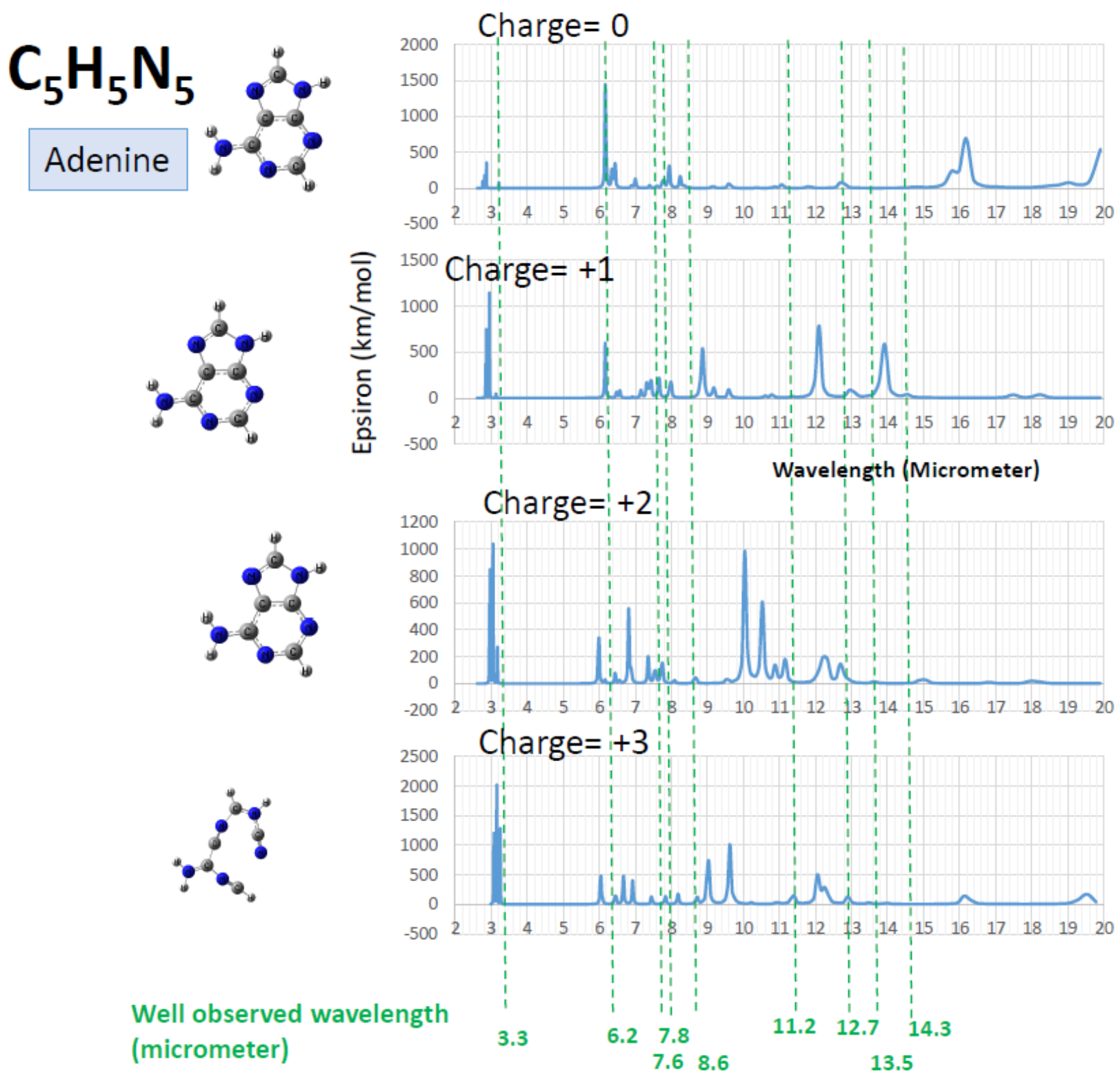

Figure 16, Molecular configuration and IR of charged adenine $C_5H_5N_5$.

Table 6, Vibrational mode analysis of purine with charge=+1. Red marked mode show not coincident with astronomically well observed one.

**$C_5H_5N_5$**
**Charge=+1**

Adenine

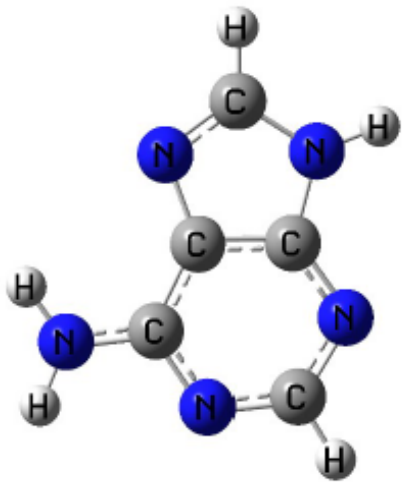

| Observed wavelength (micro meter) | Calculated IR Spectrum | | | |
|---|---|---|---|---|
| | Mode number | Wavelength (micro meter) | IR intensity (km/mol) | Vibrational mode |
| | 1 | 75.6 | 14.6 | |
| | 2 | 50.6 | 3.3 | |
| | 3 | 39.4 | 14.7 | |
| | 4 | 37.9 | 0.1 | |
| | 5 | 24.6 | 69.1 | |
| | 6 | 20.7 | 10.2 | |
| | 7 | 20.5 | 2.2 | |
| | 8 | 18.2 | 10.4 | |
| | 9 | 17.5 | 9.8 | |
| | 10 | 17.0 | 0.0 | |
| | 11 | 15.3 | 0.3 | |
| 14.3 | 12 | 14.6 | 7.9 | |
| 13.5 | 13 | 13.9 | 194.5 | N-H, C-H out of plane bending |
| 12.7 | 14 | 13.0 | 37.6 | |
| | 15 | 12.1 | 0.8 | |
| | 16 | 12.1 | 243.9 | N-H out of plane bending |
| 11.2 | 17 | 11.3 | 3.5 | |
| | 18 | 11.3 | 0.2 | |
| | 19 | 10.8 | 10.2 | |
| | 20 | 10.6 | 6.2 | |
| | 21 | 9.6 | 29.0 | |
| | 22 | 9.2 | 35.2 | |
| | 23 | 8.9 | 166.9 | C-N stretching, C-H, N-H in-plane bending |
| 8.6 | 24 | 8.8 | 39.9 | |
| 7.8 | 25 | 8.0 | 75.2 | C-N stretching, C-H, N-H in-plane bending |
| 7.6 | 26 | 7.7 | 76.5 | C-C, C-N stretching, C-H in-plane bending |
| | 27 | 7.4 | 34.9 | |
| | 28 | 7.4 | 34.9 | |
| | 29 | 7.3 | 65.6 | C-H in-plane bending |
| | 30 | 7.2 | 29.0 | |
| | 31 | 6.9 | 3.4 | |
| | 32 | 6.6 | 22.4 | |
| | 33 | 6.5 | 25.3 | |
| 6.3 | 34 | 6.2 | 175.5 | NH2 in-plane bending at NH2- |
| 3.2 | 35 | 3.2 | 0.9 | |
| | 36 | 3.1 | 21.2 | |
| | 37 | 3.0 | 376.3 | N-H stretching at NH2- |
| | 38 | 2.9 | 213.0 | N-H stretching at pentagon |
| | 39 | 2.8 | 110.7 | |

Table 7, Summary of calculated IR for three and four nitrogen substituted molecules and biological purine and adenine. Number in column was charge for every molecule. Black number means no good IR behavior compared with observed one, green number means fair. Suffix N was a case of not terminated calculation, D also serious deformation, and M deformed to mono ring from two rings. We could not find any good candidate among this group.

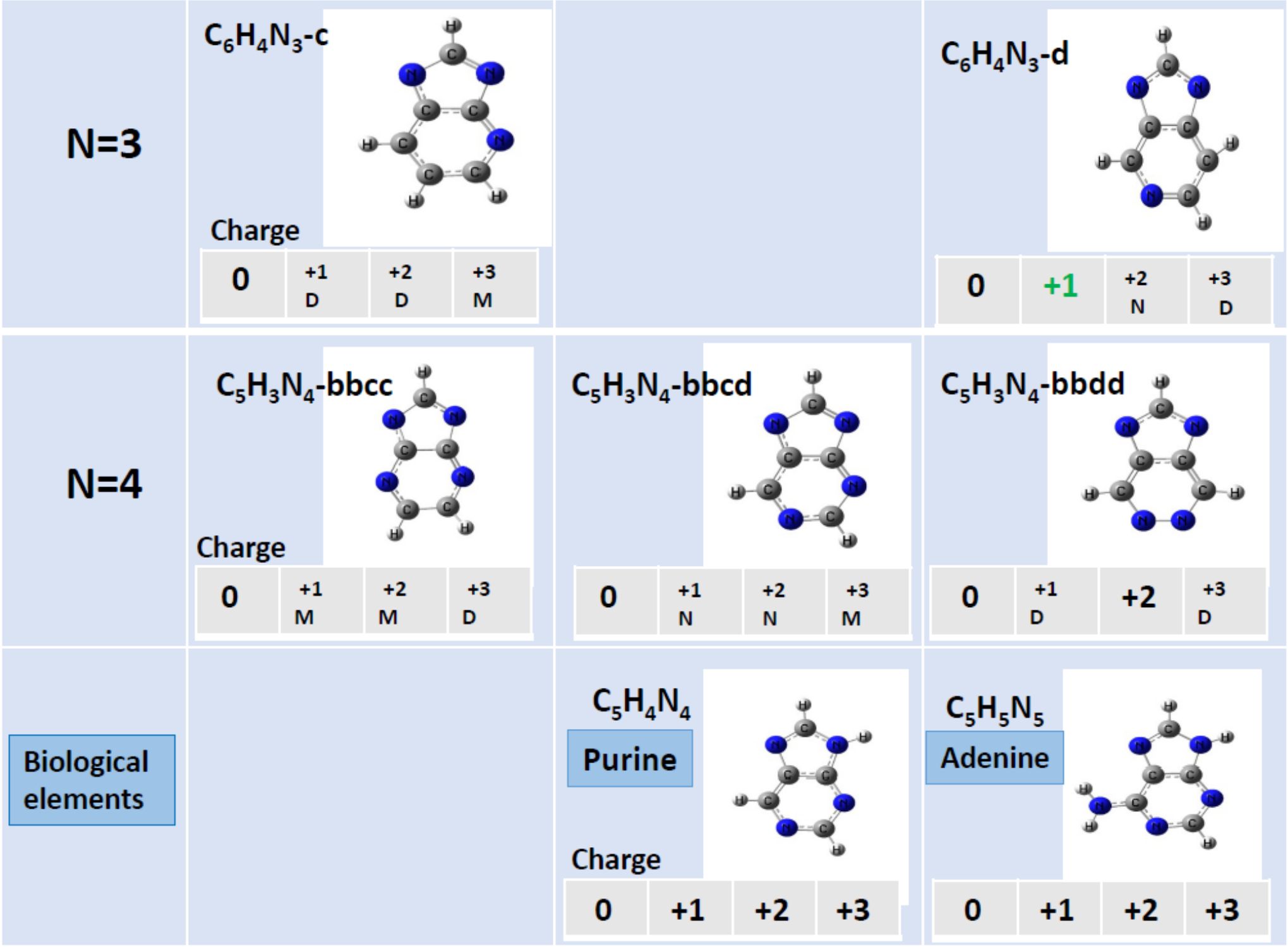

## 6, CONCLUSION

Infrared spectrum (IR) of nitrogen substituted carbon pentagon-hexagon coupled polycyclic aromatic hydrocarbon (N-PAH) was analyzed by the density functional theory.

(1) Comparing with astronomically well observed IR, one and two nitrogen substituted carbon pentagon-hexagon molecules show similar IR behavior in wavelength of 2 to 15 micrometer. Ionization was modeled from neutral to tri-cation. Among them, good candidates were $C_8H_6N_1$-b$^+$, $C_7H_5N_2$-bb$^+$, and $C_7H_5N_2$-ab$^{3+}$.

(2) IR calculation was applied to three and four nitrogen substituted ionized PAH in group of $C_6H_4N_3^{n+}$ and $C_5H_3N_4^{n+}$ (n=0, 1, 2, and 3). Calculation suggested that there are no candidates showing similar behavior. Most of them, molecules were deformed seriously by ionization.

(3) On the earth, typical biological examples of N-PAH are purine ($C_5H_4N_4$) and adenine ($C_5H_5N_5$) which are substituted by four and five nitrogen. Their calculated IR show no good similarity with observed one.

By such theoretical comparison, one capable story of chemical evolution of N-PAH in interstellar space was that one or two nitrogen substituted carbon pentagon-hexagon molecules may have a potential to be created in interstellar space, whereas more than three nitrogen substituted molecules may be not.

## ACKNOWLEDGEMENT

I would like to say great thanks to Dr. Christiaan Boersma, NASA Ames Research Center, to permit me to refer a figure (Boersma et al. 2009), also thanks his kind and useful suggestions.

## REFERENCES

Becke, A. D. 1993, J. Chem. Phys., 98, 5648

Boersma, C., Bregman, J.D. & Allamandola, L. J.. 2013, ApJ , 769, 117

Boersma, C., Bauschlicher, C. W., Ricca, A., et al. 2014, ApJ Supplement Series, 211:8

Boersma, C., Mattioda, A. L., Bauschlicher JR, C. W., et al. 2009, ApJ, 690, 1208

Frisch, M. J., Pople, J. A., & Binkley, J. S. 1984, J. Chem. Phys., 80, 3265

Frisch, M. J., Trucks, G.W., Schlegel, H. B., et al. 2009, Gaussian 09, Revision A.02 (Wallingford, CT: Gaussian, Inc.) Geballe,

Ota, N. 2014, arXiv org., 1412.0009

Ota, N. 2015a, arXiv org., 1501.01716

Ota, N. 2015b, arXiv org., 1502.01766

Ota, N. 2015c, arXiv org., 1506.05512

Ota, N. 2015d, arXiv org., 1510.07403

Ricca, A., Bauschlicher, C. W., Jr., Boersma, C ., Tielens ,A. & Allamandola, L. J. 2012, ApJ, 754, 75

Tielens, A, G, G, M 2013, Rev. Mod. Phys., 85, 1021